\documentclass[acmsmall,screen]{acmart}

\AtBeginDocument{%
  }

\usepackage{xcolor}
\usepackage{soul}
\usepackage{xurl}
\usepackage{hyperref}
\usepackage{multirow}
\usepackage{subcaption}
\usepackage[normalem]{ulem}
\usepackage{natbib}
\usepackage{graphicx}
\usepackage{multirow}
\usepackage[bb=boondox,bbscaled=.95,cal=boondoxo]{mathalfa} 

\setcopyright{rightsretained}
\copyrightyear{2025}
\acmYear{2025}
\acmDOI{XXXXXXX.XXXXXXX}

\acmConference[CSCW]{The ACM Conference on Computer-Supported Cooperative Work and Social Computing}{Oct 2025}{Bergen, Norway}
\acmISBN{978-1-4503-XXXX-X/18/06}

\acmSubmissionID{4460}

\begin{document}

\title{Trust and Friction: Negotiating How Information Flows Through Decentralized Social Media}

\author{Sohyeon Hwang}
\orcid{0000-0001-8415-7395}
\affiliation{
  \institution{Princeton University}
  \city{Princeton}
  \state{New Jersey}
  \country{USA}}

\author{Priyanka Nanayakkara}
\orcid{0000-0002-0597-6657}
\affiliation{
  \institution{Harvard University}
  \city{Cambridge}
  \state{MA}
  \country{USA}}

\author{Yan Shvartzshnaider}
\orcid{0000-0001-5954-916X}
\affiliation{
  \institution{York University}
  \city{Toronto}
  \state{Ontario}
  \country{Canada}}

\begin{abstract}
Decentralized social media protocols enable users in independent, user-hosted servers (i.e., instances) to interact with each other while they self-govern. This community-based model of social media governance opens up new opportunities for tailored decision-making about information flows---i.e., what user data is shared to whom and when---and in turn, for protecting user privacy. To better understand how community governance shapes privacy expectations on decentralized social media, we conducted a semi-structured interview with 23 users of the Fediverse, a decentralized social media network. Our findings illustrate important factors that shape a community's understandings of information flows, such as rules and proactive efforts from admins who are perceived as trustworthy. We also highlight ``governance frictions'' between communities that raise new privacy risks due to incompatibilities in values, security practices, and software. Our findings highlight the unique challenges of decentralized social media, suggest design opportunities to address frictions, and outline the role of participatory decision-making to realize the full potential of decentralization.
\end{abstract}



\received{2 July 2024}
\received[revised]{10 December 2024}

\maketitle

\section{Introduction}
\begin{figure}
  \centering
  \includegraphics[width=0.6\textwidth]{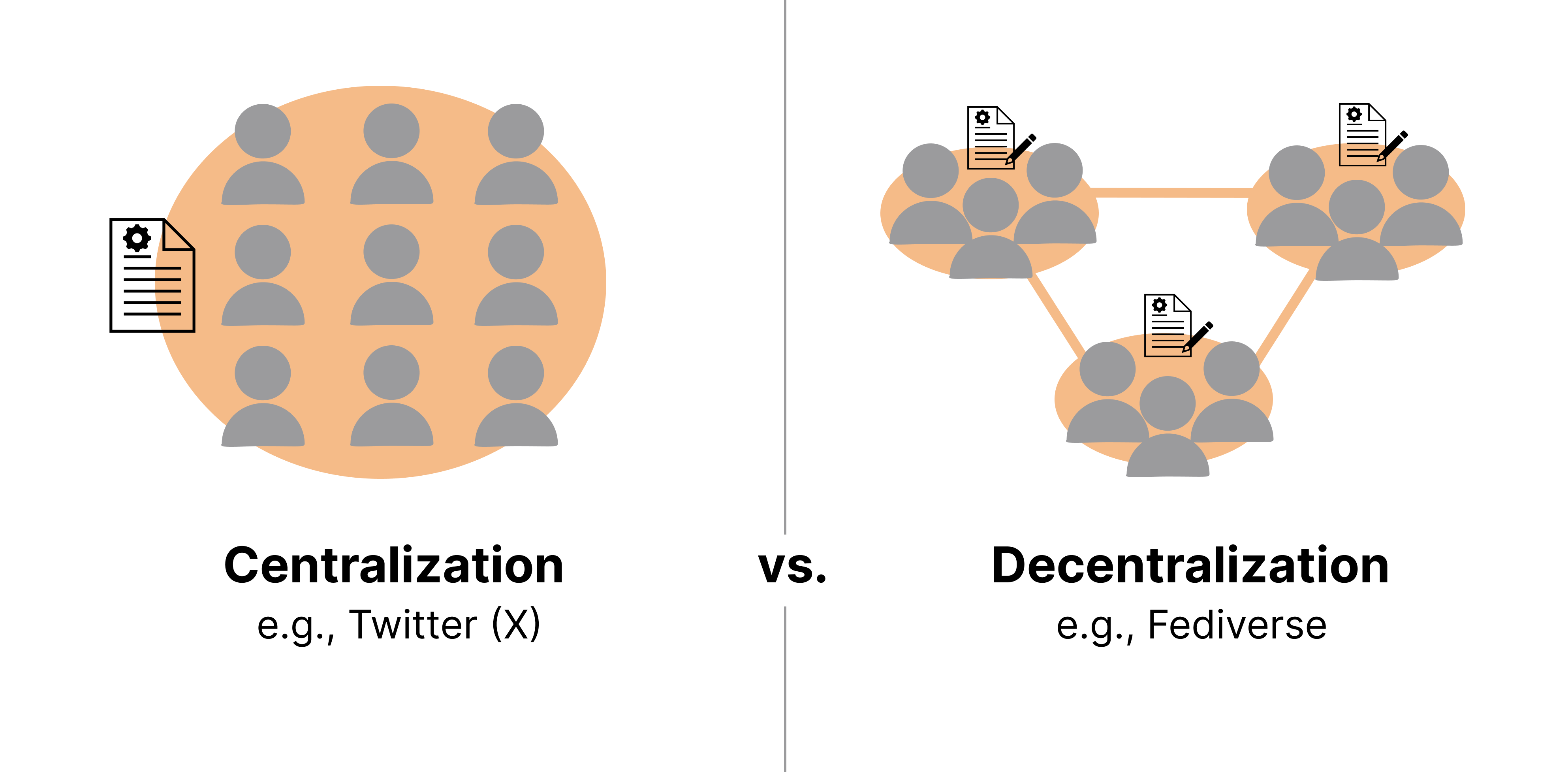}
  \caption{Users must rely on top-down decisions in centralized social media settings (e.g., Twitter). In comparison, decentralized alternatives (e.g., the Fediverse) allow users to form self-governed communities, where members can make decisions about how information flows across the system at the community level.}
  \Description{Figure showing abstractions of centralization versus decentralization in a social media settings. On the left, representing centralization, icons of people are grouped together and rules (depicted by a piece of paper with gears overlayed) appear next to them, indicating that the rules apply to everyone in the centralized platform. An example of a centralized platform is listed below (Twitter (X)). On the right, depicting decentralization, are clusters of users (each in a ``node'') in small circles. The circles are connected with lines to represent the fact that users in different communities can communicate with one another if desired. Each circle has an icon of a paper with gears on it, with an additional icon of a pencil to indicate that each community can devise their own rules for how information flows at the community level. An example of a decentralized social media setting is listed (Fediverse).}
  \label{fig:teaser}
\end{figure}

\noindent Amidst ongoing debate about centralized, privately-owned social media platforms~\citep{douekContentModerationSystems2022,odonnellHaveWeNo2021}, decentralized open-source alternatives are gaining attention  ~\citep{gehlDigitalCovenantNoncentralized2023,cohnFediverseCouldBe2022,jhaverDecentralizingPlatformPower2023,seeringReconsideringSelfmoderationRole2020}. Users on decentralized social media like the Fediverse\footnote{The Fediverse is a decentralized social media network for microblogging that has experienced rapid growth following the exodus from Twitter (now $\mathbb{X}$) in 2022~\citep[see][]{cohnFediverseCouldBe2022}.} join or set up independently-run instances (servers) that host their content and ``federate'' by connecting to a shared protocol which allows users across instances to interact. Although the interfaces visually resemble those of centralized platforms like Facebook or Twitter ($\mathbb{X}$), this decentralized design enables a form of community governance, where users (in particular, those acting as admins) can make key social and technical decisions in accordance with their values, shaping the flow of information on social media (Figure \ref{fig:teaser}). 

Because people have diverse privacy concerns~\citep{shusasAccountingPrivacyPluralism2023,soloveMeaningValuePrivacy2015,sannonPrivacyResearchMarginalized2022}, decentralization can enable groups to make more nuanced decisions~\citep{seeringReconsideringSelfmoderationRole2020} that meet their privacy needs as a community. For example, for instances using Mastodon, the most widely-used software to run instances on the Fediverse, admins can toggle on or off a series of configurations. One notable option is ``whitelist mode,'' which changes federation with other instances such that an instance will only interact with a list of approved instances --- unlike the default, wherein instances interact with each other except for those that are explicitly blocked (on blocklists). 
Instances can decide not only whether to turn whitelist mode on, but also who is included in the list and who can make decisions about it. All of these decisions shape how easily user data can be accessed by potential bad actors on other instances who might doxx or harass users, thus impacting privacy risks users must anticipate. At the same time, privacy-preserving decisions may be difficult for instances to enact, as they require active maintenance of tools like white/blocklists or routinely ensuring back-end servers (i.e, where private messages are stored) are robustly secured. Prior work has shown that community governance requires users to have 
substantial time, resources, and expertise to govern effectively~\citep{liAllThatHappening2022,zhangTroubleParadiseUnderstanding2024,hwangAdoptingThirdpartyBots2024}; these challenges could discourage many from running an instance or rely on a default but ill-suited strategy. Meanwhile, how community members can anticipate the impact of governance practices on their privacy is unclear. 
In this work, we seek to better understand how the governance practices and processes of communities shape expectations of privacy on decentralized social media, toward identifying opportunities and challenges in making privacy-preserving decisions.

We use the theory of contextual integrity~\citep{nissenbaumPrivacyContextTechnology2009} to conceptualize privacy as \textit{contextually-appropriate information flows}, i.e., whether the way user data (a user's profile, content, or other information about them) moves between actors, and when, is understood as appropriate in its social context. Drawing on 23 semi-structured interviews with community admins and members on the Fediverse, we investigate how communities self-govern to cultivate an understanding of what constitutes appropriate information flows for their particular community as part of a \textit{decentralized} social media network. We contribute a rich description of how communities manage the bounds and technical understanding of information flows among users. Our findings highlight how challenges in making privacy-preserving decisions by communities are exacerbated by ``governance frictions'' which stem from three core incompatibilities between federated communities along: values, security practices, and software. These incompatibilities highlight key challenges that will need to be addressed to advance privacy with values such as safety in mind on decentralized social media.

\section{Background}
\subsection{Privacy, agency, and safety through the lens of contextual integrity}
We follow growing theoretical consensus of privacy as a collective construct that is relational, contextual, and dynamic~\citep{marwickNetworkedPrivacyHow2014, petronioBoundariesPrivacyDialectics2002, palenUnpackingPrivacyNetworked2003, treptePrivacyCalculusContextualized2020, blackwellLGBTParentsSocial2016,viljoenRelationalTheoryData2021,nissenbaumPrivacyContextTechnology2009}, grounding our understanding of privacy in the framework of contextual integrity (CI) 
\cite{nissenbaumPrivacyContextTechnology2009}. CI defines privacy as the appropriate flow of information in a given context, with appropriateness determined by the established social norms of that context. The CI framework captures \textit{information flows} using five parameters: the data type (what kind of information is being shared), subject (who or what the information is about), sender (who is sharing the data), recipient (who is receiving the data), and transmission principle (conditions imposed on the flow). Per CI, privacy is \textit{prima facie} violated when a piece of information flows in a way that deviates from the established norms along one of these parameters, such as when information is shared with an unintended audience (recipient) or under unexpected conditions (transmission principle). 
CI has been a useful framework for making sense of privacy concerns around online participation \citep[e.g.,][]{bowserAccountingPrivacyCitizen2017,shvartzshnaiderAnalyzingPrivacyPolicies2018}, particularly in identifying \textit{when} privacy is violated and \textit{what} constitutes a violation. 
As such, the analytical advantages of CI have drawn calls in the CSCW community to leverage it in future empirical work \citep{barkhuusMismeasurementPrivacyUsing2012,kumarRoadmapApplyingContextual2024}.

Although the CI framework can help identify potentially norm-violating information flows, it does not take a normative stance on the legitimacy of an information flow beyond established norms (and not all norms are desireable to all~\citep{mcdonaldPoliticsPrivacyTheories2020}). However, recognizing that contexts are social spheres ``comprise[ed of] characteristic activities and practices, functions (or roles), aims, purposes, institutional structures, values, and action-governing norms''~\citep{nissenbaumRespectContextBenchmark2015}, CI as a framework can also bring attention to underlying structures that shape expectations about privacy in a given space. \citet{nissenbaumRespectContextBenchmark2015} details heuristics to evaluate the actors, values, and potential outcomes (benefits, harms, consequences, etc.) of contexts that information flows are a part of, in order to analyze the ethical implications of how privacy norms emerge.

A key consideration is which actors have \textit{agency} over information flows because this can change expectations about how information flows are managed. \citet{mcdonaldPowerfulPrivacyNorms2021} note that companies running centralized platforms such as Facebook, Instagram, and Snapchat exert their power over policy and design to establish privacy norms. However, when agency is shifted towards community decision-making over information flows on social media, norms about what are acceptable information flows are likely to align more closely with specific community needs. In short, user agency can radically change what is understood as an actual privacy violation, particularly because different people have different values that shape norms around privacy. For example, in the context of social media, \textit{safety} is a core value that may or may not be prioritized when determining what are contextually-appropriate flows of information, i.e., safety can shape perspectives on whether privacy has been violated (for example, did an information flow expose someone to harassment?). 
Taking the lens of CI allows us to see how key heuristics like agency and safety guide how we make normative judgments about privacy in settings like social media, where sharing information about oneself is generally ``the point.'' 

\subsection{Decentralizing social media governance to communities} \label{decentralizing}
Continued ethical and regulatory challenges surrounding popular social media platforms~\cite{douekContentModerationSystems2022,hindsItWouldnHappen2020,odonnellHaveWeNo2021} have rekindled interest in decentralized models of governing as promising ways forward~\citep[e.g.,][]{cohnFediverseCouldBe2022,jhaverDecentralizingPlatformPower2023,vincentDataLeverageFramework2021,liDimensionsDataLabor2023}. 
Recent calls for decentralization have often been framed as efforts to recoup user agency \citep{jhaverDecentralizingPlatformPower2023} over their online privacy as centralized platforms collate and sell information about users at scale and with fine granularity~\citep{boschTalesDarkSide2016,cateLimitsNoticeChoice2010,obarBiggestLieInternet2018,imLessNotMore2023,zuboffAgeSurveillanceCapitalism2019,barocasBigDataEnd2014,mcdonaldDidWatchHandmaid2023,iraniAlgorithmsSuspicionAuthentication2023}. 
\citet{viljoenRelationalTheoryData2021} argues that such datafication (i.e., the extraction and quantification of peoples' lives into data) by centralized platforms ``materializes unjust social relations'' that exacerbate inequalities. Addressing these relational harms thus requires {\em collective} institutional forms rather than just a reassertion of individual control over information, an argument echoed by related work calling for empowering bottom-up organizing around data by end users ~\citep{vincentDataLeverageFramework2021,liDimensionsDataLabor2023,dasPrivacyPeopleExploring2021,wuReasonableThingAsk2022}.

As a decentralized model of decision-making, \textit{community} governance on social media platforms offers one potential response by re-arranging the relations of data governance around groups of users. We define communities as {\em self-defined groups of users---having voluntarily and freely joined---interacting towards some common interests in a shared virtual space within a platform}, drawing on a more discrete conceptualization of ``community''~\citep{footeOnlineCommunitiesBig2023,bruckmanNewPerspectiveCommunity2006}. Because of our interest in community governance as a form of decentralized governance, we focus on communities as units of meaningful decision-making, distinguishing the concept from the ``sense of community''~\citep{blanchardExperiencedSenseVirtual2004} often associated with the term. Communities are founded by users, who govern themselves by managing membership (i.e., granting different access privileges), setting rules that guide participation, and encouraging engagement among members. Community governance refers to how a community enacts practices and processes that ``[create] the conditions for ordered rule''~\citep{stokerGovernanceTheoryFive1998} in itself.

Community governance can enhance user agency in social media by placing key decisions over information flows --- about moderation, data access, etc. --- in the hands of decentralized groups of users rather than a centralized entity. CSCW scholars have long noted how online communities can lead to more nuanced and contextually-appropriate governance decisions in large-scale social computing systems~\citep[e.g.,][]{liAllThatHappening2022,seeringReconsideringSelfmoderationRole2020,kieslerRegulatingBehaviorOnline2012}, by each representing a distinct subset of norms and goals~\citep{chandrasekharanInternetHiddenRules2018,teblunthuisNoCommunityCan2022,weldWhatMakesOnline2021}. This is particularly salient in a community's {rules}, i.e., the written normative prescriptions set by the community about how one should participate in it (sometimes also called codes of conduct, covenants, policies, or guidelines; for example, a community for sharing privacy tips might have a rule about citing credible sources). Differences in community governance can thus emerge along many dimensions, such as the processes they set for decision-making (making new rules, recruiting new admins, communicating issues, etc.), their organizational and participatory structure~\citep{schneiderGovernableSpacesDemocratic2024}, or the tools they use to facilitate governance~\citep{zhangPolicyKitBuildingGovernance2020}. In truly decentralized systems, communities may even have say over the software and technical infrastructure they are hosted on although the need for technical expertise may pose high barriers of entry.

Given the plurality of privacy needs across different socio-demographic groups \citep{shusasAccountingPrivacyPluralism2023,soloveMeaningValuePrivacy2015,sannonPrivacyResearchMarginalized2022}, communities are poised to enable more meaningful decision-making that advance the privacy interests of their particular members through community rules, processes, and tools. As communities can vary widely in purpose, views about what information flows are appropriate may similarly vary; communities can choose to enforce these views by allowing or restricting particular information flows. Understanding a community's acceptable costs and trade-offs can guide privacy-preserving governance processes and institutions~\citep{shvartzshnaiderGKCCIUnifyingFramework2022,sannonPrivacyResearchMarginalized2022}. 
For example, many communities may want have a broad, active network, such as professional communities seeking wide engagement to advance their careers \citep{wangFailedMigrationAcademic2024}. However, we can intuit that there also certain kinds of non-work content their members would not want to interact with. In general, growth is assumed to be an imperative for communities to obtain critical mass \citep{oliverTheoryCriticalMass1985} and take advantage of network effects. However, a community focused on sharing intimate, highly personal, or marginalized experiences may want their content to stay within a defined safe space~\citep{xiaoRandomMessyFunny2020,haimsonTransTimeSafety2020}. Marginalized communities especially face heightened risks of online harassment \citep{huangOpportunitiesTensionsChallenges2024,blackwellClassificationItsConsequences2017}, which has made the ability to prioritize safety in defining and maintaining privacy crucial \citep{dymVulnerableOnlineFandom2018,mcdonaldPoliticsPrivacyTheories2020}. Contemporaneous work calls for tools that strengthen ``the ability of group participants to articulate and enforce privacy norms online'' \citep{choksiPrivacyGroupsOnline2024}. 

\subsection{Shaping privacy expectations in social media communities} \label{safespaces}
A central privacy concern in any social media context is being able to anticipate the audience one is sharing information with~\citep{bernsteinQuantifyingInvisibleAudience2013,vitakImpactContextCollapse2012,vitakBalancingAudiencePrivacy2015,littImaginedAudienceSocial2016}, especially as changes in technical design can unexpectedly shift the terms of interaction among users~\citep{mulliganBridgingGapPrivacy2011,cobbUsertoUserPrivacySocial2019,triggsContextCollapseAnonymity2019,vitakImpactContextCollapse2012,marwickTweetHonestlyTweet2011}. To this end, a rich body of prior work in CSCW and related fields has examined how communities on social media can shape audience by enforcing rules around access, identity, and anonymity. 

One notable perspective in prior CSCW research looks to communities as {\em safe spaces} where people --- especially those from historically marginalized groups --- can disclose information typically considered sensitive or personal without fear of anti-social responses or offline repercussions~\citep{ammariSelfdeclaredThrowawayAccounts2019, schoenebeckSecretLifeOnline2013, dechoudhuryMentalHealthDiscourse2014, morrisonDomesticAbuseSurvivors2014, badillo-urquiolaRiskVsRestriction2019, birnholtzItWeirdStill2015,birnholtzItWeirdStill2015,guberekKeepingLowProfile2018,haimsonTransTimeSafety2020}. 
Safe spaces may focus on limiting access to the community from potential bad actors by increasing barriers to membership (e.g., requiring people apply to join the community, blocking certain users or groups \citep{geigerBotbasedCollectiveBlocklists2016}), and enforcing boundary-regulating norms more strictly. 
This work has typically focused on the experiences of individuals hoping to interact with a narrower, more sympathetic audience as they seek information, contribute their own content for others, and find new means of support online~\citep{devitoTooGayFacebook2018,kangWhyPeopleSeek2013,sannonPrivacyLiesUnderstanding2018,tranAreAnonymityseekersJust2020,huangOnionRouterUnderstanding2016}. In many cases, people participate in these spaces because they can be ``anonymous'' (i.e., dissociate content from oneself as the source~\citep{anonymousRevealNotReveal1998}), enabling them to safely share ``private sentiments in the public sphere''~\citep{asenbaumAnonymityDemocracyAbsence2018}. 

However, enabling ``anonymity'' or similar strategies is not a silver bullet solution to creating privacy-respecting and safe communities. Allowing users to be anonymous can also undermine a community's ability to provide healthy and safe interactions~\citep{sulerOnlineDisinhibitionEffect2004,grayIntersectingOppressionsOnline2012,tranAreAnonymityseekersJust2020} by creating room for harmful behaviors such as vandalism, spam, scams, toxic behavior. Cases like these warrant a certain level of information collection about individual users to enable content moderation~\citep{mcdonaldPrivacyAnonymityPerceived2019}. At the same time, such information collection can also be perceived as a kind of surveillance~\citep{bloch-wehbaContentModerationSurveillance2021,freyCanYouModerate2017,schefflerSoKContentModeration2023}, especially as it is not clear whether disallowing anonymity is necessarily effective. For example, \citet{fortePrivacyAnonymityPerceived2017} found that Wikipedians contributing via Tor --- a network which enables anonymous communication online~\citep{gorissenWhenWikipediaMet2023,mcdonaldPrivacyAnonymityPerceived2019,tranAreAnonymityseekersJust2020} --- did so out of fear of violence, harassment, reputation loss, and fear for loved ones. But because anonymous contributions on Wikipedia are associated with spam, vandalism, trolling, and abuse, Wikipedia has attempted to block contributions coming from Tor users for nearly two decades~\citep{tranAreAnonymityseekersJust2020,gorissenWhenWikipediaMet2023}. Yet, researchers have shown that Tor users that manage to bypass blocks contribute revisions of similar quality to non-Tor users~\citep{tranAreAnonymityseekersJust2020}. \citet{mcdonaldPrivacyAnonymityPerceived2019} note that community policies around contribution quality and quantity shaped whether people felt anonymous contributions should be allowed. 

These tensions demonstrate that community decisions rooted in values likes safety or quality can impact how people perceive the appropriateness of information flows. However, they are not straightforward in their effects and may require substantial iteration and deliberation. Community governance can be highly labor-intensive while voluntary~\citep{dosonoDecolonizingTacticsCollective2020,liAllThatHappening2022}: burnout is well-documented in prior research~\citep{schopke-gonzalezWhyVolunteerContent2022}. Configuring settings that have more direct implications for privacy may require technical know-how (e.g., setting up back-end servers, selecting secure services, etc.) or legal expertise (e.g., privacy policies and terms) that not all community leaders have. \citet{toschPrivacyPoliciesFediverse2024} show that, in a sample of 351 privacy policies on the Fediverse, only about 10\% had been customized, echoing broader patterns of community rule isomorphism across platforms~\citep{kieneIdentityLegitimacyVoice2024}. 

Amidst these challenges, we know precious little about which governance practices of communities shape privacy expectations, and how. In \textit{decentralized} social media, how to guide privacy-preserving governance decisions through community governance is both pressing and relatively understudied given the recent revitalization of interest in decentralization. Compared to large, privately-owned platforms --- which run on proprietary servers and code, and centrally determine and enforce a broad set of rules through design choices, algorithms, and moderation --- decentralized social media embodies critically different organizing and governing principles. Although prior research indicates that social media users rely on implicit rules (norms) that become agreed upon over time and then become taken for granted \citep{dewolfManagingPrivacyBoundaries2014,choksiPrivacyGroupsOnline2024}, how those norms become communicated and established in decentralized systems is unclear. Studies on \textit{self-disclosure} examine patterns of information sharing that provide signals of peoples' expectations of privacy \citep[e.g.,][]{wisniewskiGiveSocialNetwork2015,wangModelingSelfDisclosureSocial2016}. However, these results are primarily on centralized social media platforms and provide limited insight into how community governance might tie to privacy expectations on decentralized social media---as well as how decentralization can mitigate or produce new challenges. 

\section{Empirical setting: the Fediverse}

We examine the Fediverse, a decentralized social media network made up of independent {\em instances} (also called {\em servers}) that can communicate with each other, or ``federate,'' through a shared protocol.\footnote{At the time of writing, the ActivityPub protocol is the main protocol of the Fediverse.} A user on the Fediverse signs up on an instance (not an overarching platform), which lets them interact with other users of their instance as well as users of other instances that their instance is federated with. Often called a decentralized alternative to Twitter, the Fediverse usually consists of micro-blogging interactions like short text status updates, photos and videos, or comments to others' posts. The exact interface of the instance depends on the software it is running to connect to the Fediverse, and the most popular software used by community-run servers is a free and open-source one called Mastodon. Following our definition of ``community'' in \S \ref{decentralizing}, we use ``community'' interchangeably with ``instance'' throughout this work. 

On the Fediverse, any person can set up an instance, although technical barriers and hosting costs mean many join existing ones instead. 
Instance owners have significant governing power, shaping normative expectations for the members of their instance: what kinds of behaviors are acceptable, how content will be moderated, how data is maintained in the back-end, and which instances they will federate with. The configurations of an instance (typically, set by the admin who operates it) can shape information flows originating from and traversing its community on social media through community governance decisions. As in many online contexts, how instances make these decisions as a group varies. Some follow a ``feudal'' system where leaders act as ``benevolent dictators'' who make executive decisions~\citep{schneiderAdminsModsBenevolent2022}, while others take extra effort to cultivate community engagement in decision-making~\citep[see work supporting such efforts like][]{zhangPolicyKitBuildingGovernance2020}. Depending on how the community operates, some decisions impacting privacy --- such as how well the server's database is secured --- may not be visible or obvious to users.

The Fediverse is also notable as a home built by and for traditionally marginalized groups, in particular LGBTQ+ communities \citep{mansouxSevenThesesFediverse2020}. The ActivityPub protocol the Fediverse operates on was primarily developed by queer and trans developers \citep{karppiIFNOTELSE2023,gehlActivityPubNonStandardStandard2023}. Likewise, many conversations around safety and governance on the Fediverse have been led and advanced by these users. \texttt{\#Fediblock}, a well-known hashtag tool for tracking bad actors to block, was started by a group of queer femmes who wanted to call out a sexual harasser \citep{kiamBlacknessFediverseConversation2023}. Recent conversations have also highlighted concerns about racism on the Fediverse \citep{kiamBlacknessFediverseConversation2023,pincusAhLemmyRacism2024}. This sociocultural context is crucial in understanding privacy concerns on the Fediverse, as they are thus closely tied to questions of safety.

\section{Study design}\label{study_design}
Between October 2023---March 2024, the first author conducted semi-structured interviews via video calls with 23 users of the Fediverse. We recruited participants through a general call posted on the Fediverse and by directly contacting users. We took a statistically non-representative sampling strategy \citep{trostStatisticallyNonrepresentativeStratified1986} in recruitment, intentionally aiming for diversity in the pool along dimensions of user type (member vs. admin), user gender, server topic, and server size. Participants were offered an Amazon e-gift card valued at 20 USD. Interviews were on average 77.5 minutes long. Participants were assigned pseudo-random numbers as participant IDs (P\#s) and names were removed from transcripts. We also obfuscated details in our reporting to prevent identification. The study was approved by the [university ethics review boards].

\subsection{Interview protocol}

The interview protocol began with asking participants about their perceptions of information flows on the Fediverse, using a survey as a structured think-aloud exercise (see Appendix). Each survey item served as a vignette describing a possible information flow (drawing on CI), and participants noted the level of acceptability while explaining their choice to the interviewer. We devised these vignettes based on the language of a widely-used privacy policy on the Fediverse: in a sample of 803 servers drawn from a service listing Fediverse instances (\texttt{instances.social}), 82\% of servers used this text, which was the default privacy policy text automatically generated if an instance used the Mastodon software. Thus, we expected the policy to be representative of policies on the Fediverse broadly. 
We followed up with open questions about any privacy concerns on the instance (and the Fediverse) and reflections on the instance's practices in protecting user privacy. We also asked about how participants joined (or started) their instances for background.

\subsection{Interview pool}

\begin{table}[t]
    \centering
    \begin{tabular}{clllllll}
        \multirow{2}{*}{} 
        & & & & & \textbf{Instance}& \textbf{Instance} & \textbf{Instance}\\ 
        \textbf{P\#} & \textbf{Gender} & \textbf{Age} & \textbf{Locale} & \textbf{Role} & \textbf{software} & \textbf{topic} & \textbf{size}\\ \hline
        P43 & Male & 65+ & Western USA & Member & Mastodon & General & 1k-10k\\ \hline
        P67 & Male & 35-44 & Eastern USA & Member & Mastodon & Writing & 1k-10k\\ \hline
        P16 & Male & 25-34 & Western USA & Member & Mastodon & Tech & 10k+\\ \hline
        P22 & Male & 35-44 & Western USA & Member & Mastodon & Tech & 10k+\\ \hline
        P25 & Male & 18-24 & Midwest USA & Member & Mastodon & Academia & 501-1k\\ \hline
        P77 & Male & 35-44 & India & Admin & Mastodon & Personal & 1-10\\ \hline
        P12 & Male & 35-44 & Canada & Admin & Friendica & Personal & 1-10\\ \hline
        P61 & Male & 35-44 & Eastern USA & Member & Mastodon & Academia & 501-1k\\ \hline
        P08 & Male & 25-34 & Germany & Member & Mastodon & Academia & 1k-10k\\ \hline
        \multirow{2}{*}{} 
        & & & & & & Queer & \\ 
        P33 & Agender & 35-44 & Southern USA & Admin & Mastodon & community & 1k-10k\\ \hline
        P52 & Female & 25-34 & Midwest USA & Admin & GoToSocial & Role-play & 1-10\\ \hline
        \multirow{2}{*}{} 
        & Non-& & & & & & \\ 
        P50 & binary & 35-44 & Southern USA & Member & Mastodon & Tech & 1k-10k\\ \hline
        P73 & Female & 25-34 & Western USA & Admin & Mastodon & Subculture & 51-100\\ \hline
        \multirow{2}{*}{} 
        & Non-& & & & & & \\ 
        P11 & binary & 25-34 & Western USA & Admin & Mastodon & Subculture & 11-50\\ \hline
        P45 & Female & 35-44 & Portugal & Admin & Mastodon & Tech & 1-10\\ \hline
        P17 & Female & 35-34 & Western USA & Admin & Mastodon & Personal & 51-100\\ \hline
        P51 & Genderfluid & 18-24 & Eastern USA & Admin & Sharkey & Personal & 1-10\\ \hline
        P90 & Male & 25-34 & Western USA & Admin & Sharkey & Personal & 11-50\\ \hline
        \multirow{2}{*}{} 
        & & & & & & Queer & \\ 
        P84 & Female & 25-34 & Germany & Admin & Mastodon & community & 51-100\\ \hline
        P47 & Male & 18-24 & Canada & Admin & Mastodon & General & 1k-10k\\ \hline
        P70 & Female & 55-64 & Canada & Admin & Mastodon & Writing & 1k-10k\\ \hline
        P31 & Female & 25-35 & Eastern USA & Admin & Mastodon & Academia & 501-1k\\ \hline
        \multirow{2}{*}{} 
        & Non-& & & & & & \\ 
        P54 & binary & 55-64 & Western USA & Member & Mastodon & Tech & 1k-10k\\ \hline
    \end{tabular}
    \caption{Summary information about interview participants. Some details have been obfuscated to protect participant privacy. We replaced the instance name with a brief description of the topic.}
    \label{tab:participantpool}
\end{table}

Table \ref{tab:participantpool} shows the interview pool of 23 Fediverse users that included 11 men, 7 women, and 5 agender/non-binary/genderfluid people falling in the age range of 18-69. 
Participants were all located in the Global North (United States, Canada, Germany, and Portugal), except for one person (P77) based in India. As the majority of participants were in the United States, we aimed to gather geographic diversity within the United States; the pool includes people from 11 states.

The interview pool included 14 admins and 9 members across 19 unique instances. While most instances were running the Mastodon software, four instances used other software like Friendica. At least two more participants mentioned prior experiences with an instance that used non-Mastodon software (but not the focus of the interviews). Including these experiences expanded our analysis to examine a broader view on privacy on the Fediverse, beyond Mastodon-specific concerns.

Table \ref{tab:participantpool} shows brief descriptors of the main topic/purpose of participants' instances. Instances were general (i.e., not focused on a particular topic but on socializing on the Fediverse broadly), personal (for oneself and one's friends), and/or for a specific professional, identity-based, or subculture group. The thematically-driven groups were focused on technology issues, academia, writing, queerness, and digital media and tech subcultures (open and independent web, online roleplaying). The use of one descriptor does not necessarily preclude the relevance of others. For example, the community of P90 was mostly a personal network of friends but also considers itself a space for people of color (predominantly a group of Southeast Asians) and queer people on the Fediverse.

\subsection{Interview coding and qualitative analysis}
\begin{figure}[t]
    \centering
    \includegraphics[width=0.7\linewidth]{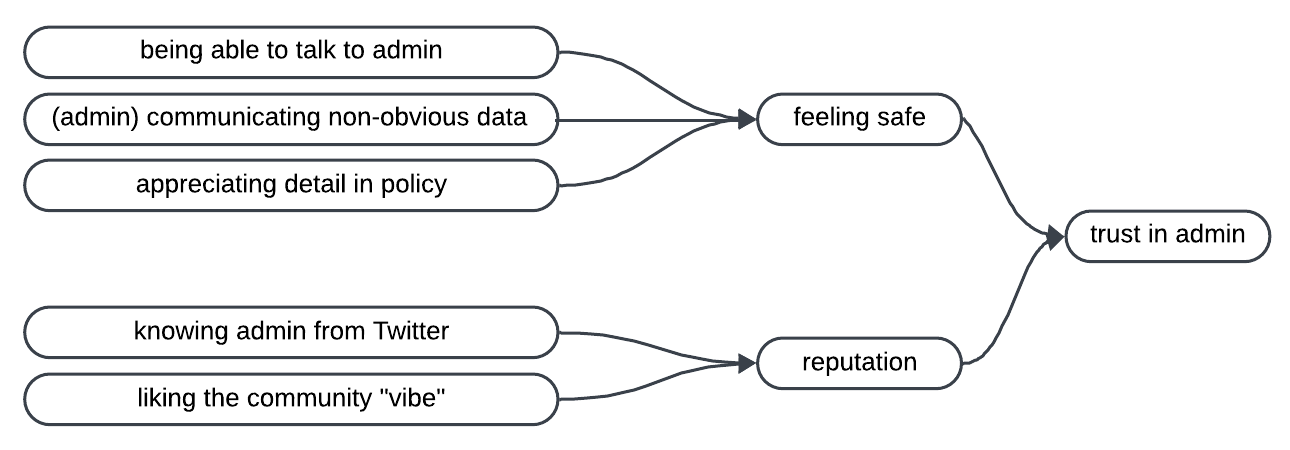}
    \caption{Example of a sample of our inductive codes (left-most) being clustered into broader themes.}
    \label{fig:qualitativecodes}
    \Description{The figure shows a flowchart of codes from qualitative analysis being clustered into broader thematic groups. Three codes ("being able to talk to admin", "admin communicating non-obvious data", "appreciating detail in policy") have arrows flowing into a thematic code "feeling safe." Two codes ("knowing admin from Twitter" and "liking the community vibe") have arrows flowing into a thematic code "reputation". The two thematic codes "feeling safe" and "reputation" have arrows flowing into an axial code "trust in admin."}
\end{figure}

Our analysis followed \citet{braunUsingThematicAnalysis2006}'s thematic analysis approach. After the first round of interviews (11), the first author summarized the interviews to identify general themes and directions in the style of memos. Following discussion with all authors, we refined the questions in the interview protocol, to balance maintaining consistency in questions asked and addressing apparent gaps in the second round (12). We determined data saturation by a heuristic of redundancy in content, themes, and codes in our interview data that suggests replicability \cite{fuschAreWeThere2015, corbinBasicsQualitativeResearch2015}.

The first author conducted line-by-line inductive coding of all transcripts after interviews were completed. This practice followed strategies recommended by \citet{charmazConstructingGroundedTheory2015}, in which initial codes closely reflect what is being described in the data, such as actions or feelings of the interviewee, and then clustered into broader axial themes (see Figure \ref{fig:qualitativecodes}). 
Through discussions with other authors who had also reviewed interview transcripts, the first author generated memos --- some based on themes identified in the first round of interviews --- that were guided by our initial motivations to understand privacy concerns in communities, and how they are addressed or not through community governance practices. Thus, our attention leaned to the dimensions of community governance participants identified as relevant to privacy. We also focused on how people were conceptualizing and reasoning about privacy in the context of their instances and decentralized social media more broadly. We developed and refined a set of findings until reaching agreement. 

\section{Establishing information flows within a community}
\label{negotiating}
We first discuss how the governance of a community shapes the bounds and technical understanding of information flows, specifically through community rules and leadership.

\subsection{Rules shape value-laden expectations around information flows}
\label{rules}
Like many online communities, instances on the Fediverse typically had a set of rules for participating in the community, such as those around harassment or content moderation. When asked about privacy risks or concerns, participants frequently referenced these instance rules. 

Participants described reviewing rules as ``\textit{philosophies that [people] have written out or borrowed from}'' (P11) to more deeply consider the shared social and political \textit{values} they conveyed. 
Rules operated as proxies of what kind of interactions would be allowed. They also signified what needs and concerns were being prioritized, especially by stating what information flows would be cut off through moderation actions (such as removing harmful content, warning users about staying respectful, and blocking instances or individuals who harassed others). To this end, the quality and detail of rules were important. P54 recalled feeling so impressed by a community's rules that they decided to sign up for an account on it. The concreteness of the instance's rules (e.g., enumerating specific behaviors that would moderated as transphobic like ``\textit{no targeted dead-naming, no targeted misgendering, cis is not a slur}'', P54) offered reassurance of what to expect:\begin{quote}
    \textit{Whether or not people have put in effort into writing server rules is an indication of whether or not this is something that the admins are seriously concerned about [...] If people are going down to that more specific level, that says to me they're probably more experienced with this kind of moderation and spend more time on it}. (P54) 
\end{quote}

\noindent Concrete, tailored rules signal care and experience in ensuring the safety of members, while vague rules could create space for bad faith interpretations. At an extreme, without rules as safeguards, bad actors could feel emboldened to ``\textit{pull up information that is not widely circulated [...] and share it freely [even if] just the amplification of that information could be dangerous [to someone]}'' (P77).

Rules mattered because they offered value-laden \textit{justifications} (or what P22 called ``\textit{legitimate use cases}'') for information flows, which impacted users' perceptions of whether they found it acceptable for an instance to handle that data in particular ways. For example, participants perceived information flows such as the collection of IP addresses for moderation purposes to be relatively more acceptable than the collection of IP addresses without any stated reason, as they felt moderation was necessary if they wanted to have functional, working social media. 

Rules also reflected a range of community goals. For example, participants in profession-oriented instances like those about tech-related careers (e.g., the instances of P16, P22, P50) and academia (e.g., the instances of P25, P61, P08, P31) noted that the point of joining the community was to increase the exposure of their community members to relevant people and content as widely as possible. Meanwhile, some communities aimed to heavily constrain the potential reach of their content, drawing stricter lines to proactively defend against potential harassment. 


\subsection{Informal and formal strategies for revising community members’ technical understandings of information flows}
\label{understanding}

Several participants felt that the Fediverse's underpinning technology --- specifically, the ActivityPub protocol --- did not prioritize privacy. Even if the software an instance used had different privacy options, interviews pointed to persistent misunderstandings around these options and how information flows on the Fediverse more broadly:
\begin{quote}
\textit{People make posts every now and again that [make me go:] Hey, listen, you know, all these different posting settings and all these different post privacy settings [that we have] are basically hacks on top of the ActivityPub protocol [...] So just be safe out there. People make posts [that reveal] quite a bit, especially when there's like big waves coming from Twitter or elsewhere.} --- P73
\end{quote}

\noindent 
The majority of our participants described themselves as having the knowledge and technical expertise to understand the Fediverse because of their professional and educational background, which allowed them to anticipate and identify violations of their privacy. As indicated by the quote above, they also felt that the average user would struggle to do the same, particularly newcomers --- most of whom were coming from Twitter and unfamiliar with the decentralized, federated model of the Fediverse. To avoid inappropriate uses and misaligned expectations, P33 argued that the most important step in protecting privacy was therefore ``\textit{educating people, because [federation] is a different paradigm [...] from other social media we're used to}.'' As the Fediverse lacked the centralized overhead that mainstream social media platforms had, the task of ``educating'' fell to the instance level via various informal and formal strategies.

\subsubsection{Informal strategies}
Participants noted that the Fediverse had a high volume of self-referential and meta-discussions. As, P54 jokingly remarked, the Fediverse was ``{\em a great place to talk about decentralized social [media]!}'' The popularity of talking \textit{about} the Fediverse on the Fediverse meant that people could develop an awareness of common privacy concerns simply by being on the network, following specific hashtags and content from admins. 

Admins in our interviews mentioned maintaining a separate account for posting updates and announcements to their community. 
For example, in the wake of corporate-owned social platforms (like Threads) attempting to add bridges to the ActivityPub protocol, P90 described posting an announcement to his community about his decision to preemptively defederate from these instances to prevent the community's content and information from being collected by corporate actors. Some admins described actively checking in with instance members about potential issues. For example, P73 reached out to users to explain some of the finer technical details about their instance (and the Fediverse in general) if users posted something that seemed rather sensitive for a general audience. Similarly, P70 described how her instance made extra efforts during newcomer onboarding:
\begin{quote}
    {\em We've certainly got a few users who have spoken to us before joining and said: I have concerns with my public account.} Can you support this? {\em And we're able to say:} Well, yes, we can --- here's the settings you should use. {\em So we're a little more rigorous about onboarding and explaining to people how they might be vulnerable and what they can do about it.} --- P70
\end{quote}

\noindent P70 posited that such extra effort was not necessarily common, particularly for instances operating at large-scale like \texttt{mastodon.social} or with open registrations (enabling user account sign-ups that didn't require any approvals). Nevertheless, the admins in our interview pool (mostly with instances of under ten thousand users) all expressed willingness to engage with users' questions. 

\subsubsection{Formal strategies}
As part of more formal ways of shaping technical understandings in the community, participants suggested that a community's privacy policy could be useful in describing how user information got collected, shared, and stored on the instance and beyond. 
Functioning as a kind of information sheet, a community's privacy policy gave instances opportunities to flag potential privacy risks, such as the fact that direct messages could be viewed by admins or what country's laws the server was be subject to. As a simple example, P33 recalled that they had specifically ``{\em bold[ed] the line that if people get messages, they can make copies of them 
}'' to warn people not to share sensitive information. 
P22 saw their privacy policy as making a plea to users to be careful before posting anything personal, especially the personal data of anyone else: ``{\em This is just, like, your dad telling you to make good decisions, right?}''

A privacy policy could also be a message to admins and users of other communities that the instance was operating in good faith, demonstrating a degree of ``{\em professionalism in running a service}'' (P47) by outlining relevant processes, technical constraints, and local laws it would be beholden to. P17 described the policy as a written ``{\em commitment in some way or another, that you aren't going to be unnecessarily endangering people who use the instance, whether it's you snooping around or doing something that allows others to snoop around.}'' 

However, privacy policies seemed to have limited impact because people do not read privacy policies \citep{obarBiggestLieInternet2018, cateLimitsNoticeChoice2010}. This pattern was evident in our interviews: admins in our interview pool noted that they rarely customized the privacy policy (at times, not realizing they had one), while members conceded that they did not read the policy beyond skimming it at sign-up. Overwhelmingly, participants described their privacy policies as boilerplate. The lack of engagement seemed to render privacy policies to exercises in compliance and legal ``{\em hygiene}''~(P54) --- even as interviewees noted that they may help convey important knowledge. 

\subsection{Leadership builds trust that information flows will adhere to expectations}
\label{admins}
We explore how community governance created a sense of trust that information flows traversing the community will adhere to expectations. Specifically, our interviews emphasized the role of community leadership, tying trust in the community to trust in the admin.

\subsubsection{Taking risks in trusting an admin}
Interviews described the inherent risks in trusting an instance admin. Participants noted that admins had back-end access to all the information people inputted as users, from sign-up information to direct message history. This access meant admins needed ``{\em to be very judicious with who they share information with}'' (P12), including law enforcement, third-party service providers, and other instances. Participants like P17 also felt admins should be mindful of how and when admins themselves accessed back-end information.

P11 acknowledged that such access was important to what admins did to keep the server running: ``\textit{I trust them to be able to understand when it would be appropriate and wouldn't be inappropriate to review that information}.''  
However, some participants reported cases in which they questioned the admin's conduct. 
In one such case, P47 recalled that the admin of their instance had enrolled all users to a third-party marketing service without their explicit consent, to email out donation requests. P47 felt their trust had been betrayed, particularly because the marketing service's design had made the unsubscribe button difficult for them to find (using black text on a black background). Ultimately they left the instance to join a different one. 

Another important consideration was whether an admin had sufficient technical expertise and knowledge to avoid security issues while maintaining the server. 
P16 appreciated that the admin of their instance ``had been operating the server for quite a while'' and was a ``really technically competent person.'' He had seen how the admin was able to adjust the server, scale it, and keep it up to date. As admins were \textit{de facto} fiduciaries of community members' data, good decision-making in the back-end ``\textit{that wouldn't compromise users too much}'' (P50) was crucial. Several participants referenced an infamous incident in 2023 where the admin of a leftist instance had kept the instance's data in a server in their home, unencrypted. When the United States federal government raided this admin's home, the server and all of the unencrypted messages between instance members on it were seized. Such incidents raised concerns among our participants that instances could serve as ``honey pots'' of users' data that could be breached or subpoenaed. 

\subsubsection{Trusting the reputation and incentives of admins}
The majority of participants expressed high levels of trust in the admin of the instance they were currently on. We identified a few factors which cultivated trust in the admin. 

Several participants noted that they knew the admin personally or through their network (or if they were an admin, that they knew their community members personally). P17, who was an instance owner, recalled that ``{\em [people joined the instance] just from me telling my friends [...] and it has just sort of steadily gone from there.}'' 
On the other side, P22 recalled seeing well-known users they liked from Twitter join the instance that he was considering: 
\begin{quote}
    \textit{Getting that kind of endorsement from [User A] and [B] and from everybody else saying that:} Yeah, we know this guy is great; he's super friendly, very level-headed, a very reasonable person. He's got a server that's been running since 2017 or something like that. {\em So there was also some history there to refer back to. And then taking a look [at the instance] for myself and being like, alright, this looks - this isn't a cesspit.} 
\end{quote}

\noindent As suggested by P22, participants went through the admin's public profile and scrolling through the recent posts on the instance as a way to gauge their trustworthiness further.  

Another important source of trust came from having well-configured, informative documentation about the server including (1) a privacy policy (even if generic), (2) lists of who was on the admin team, and (3) community rules. The combination of these showed that one had ``{\em covered all their bases}'' (P47). For P84, having such information offered transparency: ``{\em I trust my friends, and I give them access to the server. Because they are listed as moderators or administrators, it's transparent [to others] that those two or three people have access to the server}.

Some participants trusted their admin simply because the instance was not corporate-run and instead, voluntary and community-run. For example, P43 noted that although he had no idea who their admin was (``\textit{I have no reason to trust that person, and I have no reason to distrust that person.}''), he felt the admin seemed reliable. When probed further, participants emphasized the lack of financial incentives behind instances as a key part of their preconceived trust, usually placed in contrast to corporate social media that people saw as ``{\em using the data for their own good and not necessarily for the users}'' (P08). 
P43 repeatedly compared his instance to Twitter, while other participants like P25 said that they trusted their instance precisely because it ``{\em feels less corporate}.'' 
As an admin of a larger, fairly active instance, P33 addressed questions of profit directly:
\begin{quote}
    {\em We do get donations for the instance. I started collecting donations last November because the cost [of running the server] has gotten tremendously high, [...] 
    about \$1,000 a month [...] 
    It is fully paid for [by donations]. So I really appreciate the community for that. But there's not really --- there's not a profit incentive here.}
\end{quote}
\noindent As financial considerations were about covering costs rather than profit, P22 likened an instance to public radio: ``{\em[...] no one is trying to wring every dollar out of it, and it's more that we need to get the lights on and pay the salaries.}'' 
Moreover, the fairly small size of any given instance compared to centralized platforms suggested that an instance couldn't make a profit from selling data even if they wanted to. This became salient when participants described ongoing debates about whether instances should federate with emerging services developed by companies (primarily Threads) once they became compatible with the Fediverse
, as these services could re-aggregate data in corporate hands. 
Finally, participants hypothesized that constraints on the admin bandwidth made it unlikely that an admin would abuse access for personal reasons. P31, an admin herself, noted that although it was true admins could look at all data in their back-end, doing so was time-consuming and required manual searches: ``{\em The majority of us just don't care about that enough.}'' Given that other aspects of being an admin could require a great deal of volunteered time, admins emphasized a lack of interest and incentives to do extra work. 

\subsubsection{Having direct access to the admin}
Participants appreciated that they could identify a specific admin for an instance and reach out to them. 
P84 explained why the ability to connect to specific admins mattered: ``{\em It's not like we are some anonymous entities on this big website. It's more like, I can somehow relate to you. And, it's more like a friendship and not like a business communication.}'' These positive interactions with an admin as an individual and fellow user gave people the sense that the admin would operate the instance in good faith, reliably, and responsively. For example, P50 believed that their admin would reach out to them if there was a situation that might require their information to be shared somewhere, such as responding to a moderation report. P50 also emphasized that this trust had been cultivated over time, having seen the specific admin be proactive in communicating with the community. In a similar vein, participants also felt that they could hold admins accountable if needed. P22 believed that the community on his instance --- focused on technology issues --- would make sure the admin would not put the community at risk:
\begin{quote}
    \em{The guy who runs the server, we make sure that he's on top of it. I mean, he's on top of it anyway. But like, if he weren't, we would notice, and it would be --- you know, we deal with it. [...] The right people are paying attention, I suppose.}
\end{quote}

\section{Governance frictions between communities}
\label{friction}

As communities made governance decisions, they naturally differed in how they sought to shape information flows. Differences could lead to \textit{governance frictions} between communities: incompatibilities between communities that, on one hand, could enforce the goals of one community but on the other hand, potentially undermine the goals of the other. In the context of privacy concerns, governance frictions could pose new risks to privacy in a community or limit the effectiveness of a community's strategies to protect the privacy of its users. We identified three main incompatibilities between instances that had consequences for privacy: value, security, and software.

\subsection{Value incompatibilities} 

The majority of the incompatibilities mentioned by the participants were due to differences in ``{\em communication cultures}'', as referred to by P08, that impacted what would be acceptable content and behavior. These differences arose because communities had ``{\em different interpretations of [rules] and different levels of granularity of how far they'll go in enforcing different bits of [them]}'' (P73). 

Participants generally viewed normative differences as a positive feature, not a bug, of the Fediverse. As P73 put it: ``{\em I feel like it really makes it so everyone can find their place [...] on the Fediverse [...] that's one of the things that's really beautiful about the Fediverse to me.}'' Similarly, P54 felt that the diversity of norms across the Fediverse made it much more interesting and vibrant: ``{\em The great thing about the Fediverse is the combination of these local communities at the instance level within this broader conversation. Instances have different norms, instances have different vibes.}''

However, at times the values behind these norms differed enough that people did not want to interact with parts of the Fediverse. For example, P73 noted that some instances did not want to see nudity or other NSFW content because it clashed with their goals for being on the Fediverse (such as professional branding and posting). Because values were social and political, the stakes could be high. P45 observed that the queer members of their community had concerns about being outed (as part of the LGBTQIA+ community) or ``{\em being harassed or doxxed or whatever [for their views]},'' as queer people are disproportionate targets of online harassment~\citep{blackwellClassificationItsConsequences2017}. Meanwhile, some instances embraced ``{\em free speech absoluti[sm]}'' (P73) and allowed abusive content in the name of free speech rights. These differences raised questions of how federation could potentially put one's instance members' reputations at risk as well as put them in harm's way of abusive behavior. As P45 stressed: ``{\em The main thing that I think about when it comes to privacy concerns [on social media] is that you don't want to share stuff that makes you feel vulnerable.}''

Defederating (and blocking more generally) enabled communities to address feelings of vulnerability by cutting information flows between themselves and value-incompatible instances. Participants perceived blocking access to one's information (profile information, posts, online activity) or blocking ``{\em people who are harassing you [...] or attempting to surveil you}'' (P52) to be ``{\em intimately linked with privacy}'' (P52). Several participants who were admins described making decisions to defederate to protect their members from discriminatory behavior and content, as well as morally reprehensible content such as child sexual abuse materials (which is also illegal, putting instances at legal risk as well). P33 explained that defederating was a swift decision when it came to value incompatibilities: 
\begin{quote}
    {\em More often than not, I'm really referring to abusive behavior. You know, it's not, it's not illegal for somebody to use curse words at me and call me slurs online. But, you know, I don't want to see it. And I don't want our community to be involved in that.}
\end{quote}

\noindent However, echoing other participants, P16 noted that blocking individuals or defederating from an instance has limited effectiveness in governing the flow of information online: 
\begin{quote}
    {\em If I block them, you know, maybe they can't follow me with their regular account, but they can make another account where they can do whatever. If something's on my profile, I know it's not going to be something that I can hide from anybody on the internet --- like, it's out there. So, you know, I prefer that if I can block somebody, they just suddenly couldn't see my stuff at all. But I know, that's not reality.} 
\end{quote}

\noindent In addition, in a variant of ``doxxing,''  people could maliciously expose a person to another instance or person one had blocked. P73 and P47 recalled seeing instances that {\em intentionally} operated bots that would find out who blocked whom, and tag both users in the same post to force them back into contact with each other. In other cases, this exposure could be accidental, like if a post got shared to a different instance that didn't block the same instances as the original poster's instance. In short, being federated to many communities meant having many potential points of leakage.

To further reduce the risk of these kinds of privacy violations, some instances opted to operate in ``allow-list'' mode in which the server would only federate with approved instances (instead of the more widely-used practice, where servers defederate from instances over time) so that ``{\em random people couldn't see their [content]}'' (P52). However, participants noted that the ``allow-list'' mode is a drastic and somewhat controversial measure that could be at odds with the information flows people {\em did} expect on social media. 

Ultimately, the interview participants did not demonstrate consensus about how to mitigate privacy risks posed to users due to the value incompatibilities. However, across our interviews, participants stressed the importance of keeping up to date with meta-discussion about evolving safety issues across the Fediverse. As P12 put it: if one wanted to properly prevent information from spreading to certain spaces, there needed to be ``{\em a lot of intentional chatting}.'' 

\subsection{Security incompatibilities}
Participants described potential issues around instance security, which spanned all aspects of how an instance might choose to govern access to the instance (e.g., registration, federation) and its back-end (e.g., admin access to logs and databases, encryption of stored data, credit card info for donations to support server costs). 

Participants expressed concerns that an instance's poorly-devised security processes could undermine the security of an instance's operations. For P67, the fact that some instances could simply use ``{\em informal and social arrangements}'' could be ``{\em anxiety-inducing}.'' For example, having open registrations could let bad actors and bots into an instance, inadvertently impacting people on instances federated with the compromised instances. Most admins with whom we spoke had semi-closed registrations on their instances and often saw open registrations as a ``{\em red flag}'' (P17). As P51 suggested, poor security configurations of an instance created a weak point that could affect federated instances even with relatively strong security settings: 
\begin{quote}
    {\em Even if the admin on my instance --- AKA, me --- tries to be very security conscious and all, if the fine admin of some other instance were to make a security oopsie, a gaffe, a failure... Then that is going to affect anyone [...] who has had dealings with them. But beyond that, I don't think there's really much of a threat model that other servers pose [...] The only legitimate added threat is if private user information on my instance ends up on another instance through any number of reasons, and if it's not secure over there, that poses a risk to the users of my instance, not to that instance alone.}
\end{quote}

As with value incompatibilities, participants shared that they mainly use the ``blocking'' feature on compromised instances to avoid potential security breaches. While crude in the granularity at which it could be applied, blocking enabled participants to draw ``{\em boundary lines with the rest of the Fediverse}'' (P70). However, participants also noted that instances varied in what they considered a security threat and what configurations would be sufficient to preserve user privacy, even if they shared the same overarching community values. For example, participants pointed to the aforementioned debate about federating with corporate-run instances such as Threads. While some like P90 felt federating with Threads would be a major threat by giving data to corporate actors, others did not. They argued that one's posts ``{\em could still get sucked up [...] if your posts end up on a server that is friendly with Threads}'' (P17). As P70 observed, dealing with security differences was important but required a great deal of work as an ongoing negotiation, ``{\em even [between] adjacent administrators who are maybe on the same wavelength}.''

\subsection{Software incompatibilities} 
Instances used software for running services on the Fediverse with different technical affordances for privacy to their users. P12, for example, recalled how he and a fellow Fediverse user had realized that the technical implementation of marking messages as ``private'' differed between their instances because their instances used different software (Mastodon and Friendica):
\begin{quote}
    {\em I was following [a Mastodon user] and I boosted a post of theirs. And then they sort of screamed at me because I was boosting a private post. And I was like:} Well, then mark it private! {\em And they were like:} But I did! {\em And then we figured it out. [...] That's what I mean when I talk about it being impossible to control at that point. Like, if I reply to something [marked private on Mastodon], that post would be shown to other people who use Friendica, because that post isn't marked private [on Friendica]}
\end{quote}

\noindent By itself, the protocol underpinning the Fediverse does not define the notion of messages marked ``private.'' Instead, developers of softwares like Mastodon or Friendica can build additional functionality. As P73 noted, marking a message private was a soft ``{\em hack}'' (P73) in the Mastodon software alone that had no meaning to the Friendica software. Moreover, these softwares were usually updated over time, which meant instances needed to keep posted on software updates; in a large federated environment, it is difficult to guarantee that all instances are using the up-to-date version of the software. 
Importantly, an instance's software wasn't compelled to respect privacy options used by another instance: ``{\em [The software] isn't really a wall. It's a stop sign, you know?}'' (P51). Differences in implementations and versions of softwares revealed, as P51 continued, ``{\em an illusion of privacy}'' with  ``{\em no actual privacy at the heart of [the protocol]}.''

The technical differences across instance software described above could generate conflict between users who believed the other had carelessly violated their privacy. Nevertheless, in our interviews, relatively few participants raised this issue because most people were on an instance using the Mastodon software and primarily interacted with people on instances using the same software. Some, like P84, even stated that they wanted to see more softwares on the Fediverse because they liked ``{\em that [they] can interact with all kinds of different software [...] implementations that rely on ActivityPub, and that software works together instead of being always its own silo}.'' 

The rise of Mastodon as the dominant software of the Fediverse puts it under larger scrutiny as to whether it offers adequate privacy options to users across the many instances running it. P84 felt that moderation tools on Mastodon were insufficient, while P51 observed a sense of {\em ad hoc} standardization on the ActivityPub protocol driven primarily by wanting to federate with Mastodon instances (in particular, the flagship instance with nearly a million users, \texttt{mastodon.social}) rather than by better security or privacy practices. P90 highlighted a recent bug found on Mastodon, noting that instances using different software (including his own) had avoided the problem: 
\begin{quote}
    {\em It was basically something like, you could send a fake post and it would overwrite the real one or someone's profile. That's a pretty serious vulnerability. You could impersonate anybody, and apparently it was really easy to do. But Sharkey was not vulnerable. So we didn't have that vulnerability even to begin with.} 
\end{quote}

\noindent Where technical differences in softwares could pose a challenge to privacy in earlier examples, P90's story suggested that technical differences could also preserve user privacy by limiting the impact of vulnerabilities from one particular software. In sum, software incompatibilities could lead to significant privacy-related harms but participants noted that there was ongoing debate among Fediverse users about how to resolve these issues. Participants suggested that improving the core privacy-functionalities of the ActivityPub protocol instead of through software like Mastodon or Friendica could provide a consistent way forward for privacy-preserving federation.

\section{Discussion}
\label{discussion}

\subsection{The unique challenges of decentralization in social media communities}
Echoing work in other social media contexts \citep[e.g.,][]{fortePrivacyAnonymityPerceived2017,dymVulnerableOnlineFandom2018,scheuermanSafeSpacesSafe2018}, our interviews emphasized safety as an important consideration when governing privacy at the community level, particularly in discussing community rules (\S \ref{rules}). However, the varying goals of communities meant that they had different ideas of what constituted safety. In traditional online communities, normative difference is expected \citep[see][]{fieslerRedditRulesCharacterizing2018,chandrasekharanInternetHiddenRules2018} and can even be an advantage~\citep{teblunthuisNoCommunityCan2022}. On community-based platforms like Reddit, prior work notes that despite such difference, 
inter-community conflict appears to be relatively rare~\citep{kumarCommunityInteractionConflict2018}. However, on the Fediverse, communities are \textit{federated}: creating and sharing content with one another, inducing information flows that may or may not be appropriate for each of their social contexts. As evidenced by the governance frictions in \S \ref{friction}, the increase in potential cross-community information flows produces a unique challenge for decentralized social media in defining and enforcing safety in privacy decisions.

One admin noted they were highly selective about other communities they federated with, and a few suggested that this strategy could ensure a sense of safety. However, participants also noted that this approach was in tension with the general purpose of the Fediverse: to be a social media network, possibly one that could be as connected and influential as centralized social media platforms. Our participants represented communities with a diverse range of goals. As such, some communities saw benefits from broader networks (e.g., professional, research, hobbies) while others wanted to limit reach and access (e.g., subcultures, queer communities, personal groups). Hence, future work will need to identify ways to implement safety in accordance with the different breadths of networked reach that communities want, whether by tools, policies, or new protocols.

Users' past experiences on centralized social media also exacerbated another problem on the Fediverse: newcomer onboarding. As noted in \S \ref{understanding}, many communities found that new members were coming from Twitter with a strong but misaligned mental model of how social media ``works.'' 
Admin efforts were thus drawn to correcting misunderstandings about the technical inner workings of the Fediverse. As noted in our description of security and software incompatibilities, correcting misunderstandings was especially important because technical differences between communities could produce disjunctions in features intended to protect privacy. Newcomer onboarding is a classic problem in the development of any social computing system \citep{krautBuildingSuccessfulOnline2012,burkeFeedMeMotivating2009,morganTeaSympathyCrafting2013}. Focused on privacy as the guiding heuristic, our work underscores that onboarding practices in decentralized social media must pay further attention to ensuring newcomers understand the technical and infrastructural aspects of each community: where the server is hosted, who has backend access to data, how information is shared between servers, and so on. 
Future work might draw on the rich body of prior research building interactive tutorials \citep{narayanWikipediaAdventureField2017}, testing new social spaces for newcomers \citep{morganTeaSympathyCrafting2013,warncke-wangIncreasingParticipationPeer2023}, and strategically placing rules \citep{matiasPreventingHarassmentIncreasing2019} to strengthen and improve the onboarding process.

\subsection{Designing to mitigate governance frictions}
We identified three kinds of incompatibilities that produce governance frictions for privacy in decentralized social media. Designing tools and interfaces that help communities better respond to and anticipate governance frictions may help improve users' experiences in the Fediverse, especially given how these frictions are how problems emerge. Participants noted that the main tool available to them was a variety of blocklists, which enabled them to defederate from communities \textit{en masse}. Blocklists, while an important option to have, are primarily about value incompatibilities that become salient in content moderation practices and rules. As a result, security and software incompatibilities remain mostly unaddressed. As each incompatibility will require different tooling and design to resolve, future work should avoid conflating the three. 

Our interviews emphasized a lack of consensus about how to deal with incompatibilities alongside a pressing need to resolve them. In \S \ref{negotiating}, our findings show how communities handled potential confusion about privacy (including issues that emerge due to incompatibilities) \textit{within} the community through consistent communication, such as an admin account posting updates. We suggest that, likewise, designing opportunities for signals, communication, and negotiation \textit{between} communities can help address the kinds of incompatibilities leading to governance frictions shown in our work.
Below we briefly provide an example of a potential approach per incompatibility:

\begin{itemize}
    \item \textbf{Value incompatibilities}: While participants look to written rules to get a sense of a community's values, differences in rules do not always equate value incompatibilities. However, recent work shows that communities often share or have similar rules that they copy from reliable sources to signal legitimacy \citep{kieneIdentityLegitimacyVoice2024}. Thus when differences are significant, warning flags during moments of federation might prompt admins or members to review or open dialogue with another community. This might be done through natural language processing (NLP) methods to evaluate similarity or contradiction between rule sets.
    \item \textbf{Security incompatibilities}: Communities might develop ``standards'' for levels of security an instance implements. This could be communicated clearly with badges that indicates a certain set of criteria related to the instance's security are met. Badges can enable other instances to make quicker and more effective decisions about whether an instance's standards are up to par. How these standards are developed is an open question and suggests it may be good to have a mechanism for admins or representatives from different instances to convene, discuss, and later change Fediverse-wide decisions as a collective. 
    \item \textbf{Software incompatibilities}: Our interviews noted that the software used to run instances on the Fediverse were inconsistent in how they treated posts as ``private'' (i.e., which audience the post will be shown to). We suggest developing concise abstractions that convey such incompatibilities. For example, widgets might automatically show, based on an instance's software, where a certain post will travel to. Given that this should be consistent to be useful, focusing on new features or capabilities at the protocol-level will likely be a helpful first step. As with security practices, how this idea might be developed as a standard in the protocol is an open question and will require coordination and input from diverse communities.
\end{itemize}

These suggestions emphasize encouraging interaction between communities. In many ways, this contrasts with defederation, which is the main mode of dealing with conflicts or incompatibilities on the Fediverse. While defederation is a crucial feature for addressing online harm \citep{colglazierEffectsGroupSanctions2024}, our interviews indicated that there were many simple workarounds that could be leveraged by malicious actors. Meanwhile, security and software incompatibilities less frequently warrant defederation but still must be resolved. Strategies that complement defederation as an option available to communities are crucial in enabling communities to strengthen privacy on decentralized social media systems like the Fediverse. Our suggestions above are just a few potential paths moving forward.

To this end, because incompatibilities are fundamentally about information flows, returning to the framework of contextual integrity can prompt normative evaluation to generate recommendations and design ideas. \citet{choksiPrivacyGroupsOnline2024} identifies social roles in online groups and places them in information flows as actors to analyze privacy challenges within groups. A similar exercise evaluating information flows across groups could be useful to guide design in surfacing incompatibilities or reasoning why incompatibilities raise privacy risks. For example, developing badges for security incompatibilities helps communities better understand the transmission principles of information flows between two actors (communities), one of which may need much stricter security practices that make another's weaker practices norm-violating.


\subsection{Toward more participatory decision-making in decentralized social media}
Whether or not a user could trust their community admin was a major theme in our findings, as admins held executive power over much of how a community (i.e., instance) worked. This arrangement echoes the ``feudalistic'' forms of governance commonly found in many online communities \citep{schneiderAdminsModsBenevolent2022}, where only a few individuals have the power to make decisions and function as ``benevolent dictators.'' In {\em Governable Spaces}, \citet{schneiderGovernableSpacesDemocratic2024} critiques this pattern as cultivating undemocratic digital spaces. Recent work in CSCW has attempted to enable more participatory forms of decision-making \citep{schneiderModularPoliticsGovernance2021,lamEndUserAuditsSystem2022}. 
Our findings suggested that people trusted their admins would operate in good faith, while also recognizing the inherent risk in trusting a person they might not know. In \S \ref{admins}, we describe how trust was driven by how accessible an admin was and how proactive they were. 
This suggests that even in the absence of formal or explicit participatory structures, communities can avoid \textit{de facto} feudalism. For example, participants of one of the communities we spoke to noted that the average tech expertise of community members meant that community members were quick to call out privacy discrepancies. In other communities, admins were beholden to social or reputation costs (e.g., a server for personal friends), which motivated them to engage in participatory decision-making informally. 
To move toward participatory decision-making, future work must systematically evaluate the conditions under which having an admin provides an advantage for community outcomes --- and what mechanisms help safeguard a community from a ``dictator.'' In the context of privacy, articulating these conditions can help us understand when and where more formal participatory structures of governance would strengthen the maintenance of privacy in community-based approaches to governing social media. Future work investigating how participatory decision-making subsequently shapes how issues like governance frictions are resolved will also be crucial in improving the design of decentralized systems overall. 

\subsection{Limitations}
\label{limitations}
We highlight study limitations that would benefit from further consideration in future work. Participants emphasized how membership in marginalized identity groups shifted the privacy risks people faced, especially due to racism and queerphobia. While many of our participants were part of the LGBTQIA+ community, we did not systematically collect information about race or ethnicity. Given the importance of safety in our findings, future work should explicitly center the perspectives of marginalized communities toward developing ``equitable and privacy-protective technologies''~\citep{sannonPrivacyResearchMarginalized2022}. We also note that all but one of our participants were located in the Global North, but the Fediverse has seen growing popularity in countries like Brazil. Different legal paradigms and political contexts impact privacy concerns as well as shape community practices. Future work on non-United States and non-European contexts in particular will offer valuable insights.

Our interview protocol involved discussing scenarios with CI-based vignettes (as a survey) based on a privacy policy commonly seen in a sample of Fediverse instances. However, privacy policies often omit contextual parameters that impact how people judge information flows to be appropriate~\citep{shvartzshnaiderAnalyzingPrivacyPolicies2018}, which may lead people to rely on preconceived (and inaccurate) mental models of privacy~\citep{martinPrivacyNoticesTabula2015}. We used the survey as an artifact for open discussion so the interviewer could clarify interpretations and ask follow-up questions. As we found the CI-based vignettes useful for evoking conversation about information flows on social media, future work might design and leverage CI vignettes using other guidelines that might make other privacy issues (concerns of marginalized communities, different legal paradigms or geopolitical contexts, etc.) more salient in analysis.

\section{Conclusion}
Grounded in the theory of contextual integrity, this study describes how community governance enabled participants to engage in an active negotiation of privacy expectations by shaping the bounds, trustworthiness, and technical understandings of information flows. Our findings suggest that communities develop shared expectations around information flows through concrete rules and proactive leadership. Together, these practices seem to foster trust among users that privacy expectations will be met. As such, community governance can offer valuable opportunities to re-imagine privacy on social media, particularly toward prioritizing values like safety. These opportunities do not come without hurdles. \textit{Governance frictions} could generate tensions due to incompatibilities between communities' values, security concerns, and software, which can give rise to novel risks and privacy challenges. Future work must take stock of the unique challenges decentralization brings and work toward stronger community-based governance mechanisms that support processes of negotiation around how information flows move through communities. 

\begin{acks}
We are grateful to our interview participants for their generosity and time. We would also like to thank Kaylea Champion, Aaron Shaw, Darren Gergle, and Ágnes Horvát for their feedback, reviewers for their constructive comments, as well as attendees of the 9th Science of Community dialogue and the 2023 Symposium on Contextual Integrity for their valuable input. 

We acknowledge the support of the Natural Sciences and Engineering Research Council of Canada (NSERC), RGPIN-2022-04595, the National Science Foundation (DGE-2234667, IIS-1910202) and the Harvard Center for Research on Computation and Society.
\end{acks}

\bibliographystyle{include/ACM-Reference-Format}
\bibliography{references}

\appendix

\section{Research Methods}

\subsection{Interview Protocol - Survey Exercise}

As noted in \S\ref{study_design}, the first part of our interviews involved using a survey as an artifact for structured discussion. The interviewer shared their screen showing the survey and walked through survey items with participants, who said their choices aloud and then explained their reasons for their choice (at times followed by further prompting or clarification questions by the interviewer). Each item (bullet dash) on the survey had the following options: ``Not at all acceptable,'' ``Somewhat acceptable,'' ``Acceptable,'' ``Completely acceptable,'' and ``I do not wish to answer this question.'' Items were arranged under two blocks, as follows:

\begin{itemize}
    \item Rate the extent to which you find it acceptable that an instance collects and uses data (e.g., personal information someone has entered, posts, DMs and comments someone toots on the server) for each of the following reasons.
    \begin{itemize}
        \item To share your posts and activity with other communities or users
        \item To send you information, notifications about other people interacting with your content or sending you messages
        \item To automatically save your preferences for future visits with cookies
        \item To aid moderation of the community, like comparing your IP address with known banned IP addresses
        \item To retain server logs, like all IP addresses
        \item To share with trusted third parties who assist the server in operating the site, conducting business, and servicing users, so long as those parties agree to keep this information confidential
        \item To comply with the law, enforce site policies, or protect the server's or others rights, property, or safety
    \end{itemize}
    \item Rate the extent to which you find it acceptable for each potential type of person to come across your public information, posts, and comments originally shared on your instance.
    \begin{itemize}
        \item Someone you DM or @
        \item Your followers and/or people you follow
        \item Someone you've blocked on the Fediverse or elsewhere
        \item A person on the Fediverse you have no connections to
        \item Admins, moderators, and other operators of instances (yours and/or others)
        \item An app collecting online data generally
    \end{itemize}
\end{itemize}

\end{document}